\documentclass{article}
\usepackage{geometry}
\usepackage{authblk}

\usepackage[utf8]{inputenc}
\usepackage{graphicx}
\usepackage{longtable}
\usepackage{caption}
\usepackage{subcaption}
\usepackage{booktabs}
\usepackage{xtab}%
\usepackage{afterpage}
\usepackage{colortbl}%
  \newcommand{\myrowcolour}{\rowcolor[gray]{0.97}}

\usepackage{xcolor}
\usepackage{natbib}
\usepackage{pdflscape}

\title{Publishing patterns reflect\\ political polarization in news media}

\author[1]{Nick Hagar}
\author[2,3]{Johannes Wachs}
\author[1]{Em\H{o}ke-\'Agnes Horv\'at\thanks{Direct correspondence to: nicholashagar2018@u.northwestern.edu \& a-horvat@northwestern.edu. This work was partially supported by the U.S. National Science Foundation under Grant No. IIS-1755873.}}

\affil[1]{ Communication Studies, Northwestern University}
\affil[2]{ Vienna University of Economics and Business}
\affil[3]{ Complexity Science Hub Vienna}
\date{January 2021}

\begin{document}

\maketitle

\begin{abstract}
Digital news outlets rely on a variety of outside contributors, from freelance journalists, to political commentators, to executives and politicians. These external dependencies create a network among news outlets, traced along the contributors they share. Using connections between outlets, we demonstrate how contributors' publishing trajectories tend to align with outlet political leanings. We also show how polarized clustering of outlets translates to differences in the topics of news covered and the style and tone of articles published. In addition, we demonstrate how contributors who cross partisan divides tend to focus on less explicitly political topics. This work addresses an important gap in the media polarization literature, by highlighting how structural factors on the production side of news media create an ecosystem shaped by political leanings, independent of the priorities of any one person or organization.
\end{abstract}

\section{Introduction}
Political polarization in digital news consumption is well established. While the existence of self-reinforcing isolation that is strong enough to \textit{override} individual agency---so called ``filter bubbles''---is hotly contested, there is substantial evidence of patterns of media consumption aligned with political identity in the U.S. \citep{bakshyExposureIdeologicallyDiverse2015a,macyOpinionCascadesUnpredictability2019,shiMillionsOnlineBook2017}. Whether by an explicit recognition of political identity or the implicit association of certain behaviors and traits with a politically-aligned community, in-group identity drives members of opposing political groups into silos \citep{colemanSocialCapitalCreation1988,dvir-gvirsmanMediaAudienceHomophily2017}. However, this work only explores consumption dynamics.

As a result, researchers have often sought to test the efficacy of cross-cutting content and conversation in reducing polarization, which exposes members of opposing groups to challenging viewpoints \citep{heatherlyFilteringOutOther2017}. The idea of intentional exposure to conflicting views finds its way into proposed interventions. Past work on ``bursting'' filter bubbles introduced tools that allow users to monitor the political polarity of their news consumption, for example, or to seek out the opposing views on important issues \citep{flaxmanFilterBubblesEcho2016,resnickBurstingYourFilter2013}. In an effort to support consumers, other work imagines alternative modes of suggesting media to users via recommendation systems \citep{helbergerDemocraticRoleNews2019}. These interventions address exposure to news, but not its production.

In news production, research often focuses on the individual priorities and professional norms countervailing to polarization. In particular, researchers pay close attention to the professional norm of objectivity within journalism and how it gets enacted to minimize partisan influence \citep{schudsonObjectivityProfessionalismTruth2009,mcchesneyProblemJournalismPolitical2003}. This perspective focuses on the individual journalist as the counterpoint to polarization in news media, as they carefully construct coverage with an eye toward fairness, balance, and accuracy \citep{ryanJournalisticEthicsObjectivity2001}. And while this view acknowledges the role of news producers in polarized environments, it also fails to address the structural forces shaping that production.

Research that does take a structural approaches finds ample evidence of polarization in the distribution of digital news. The earliest political blogs demonstrated an overwhelming preference for linking to and engaging in dialogue with like-minded outlets \citep{adamicPoliticalBlogosphere20042005}. Similarly, work on the dissemination of digital election coverage found an isolated, self-reinforcing group of right-leaning outlets, one that largely disengaged itself from the rest of the news ecosystem \citep{benklerNetworkPropagandaManipulation2018}. Third-party social media platforms used for news distribution seem to largely amplify these producer preferences \citep{bakshyExposureIdeologicallyDiverse2015a,benklerNetworkPropagandaManipulation2018,wihbeySocialSilosJournalism2019}. Therefore, even in a world where media consumers have the agency to cross political boundaries and discover opposing views, the distribution of that media operates in such a way that discourages cross-cutting exposure. Polarization is an integrated problem.

This study attempts to extend that structural view into the production of digital news itself, by examining the movement patterns of news contributors from outlet to outlet. We treat these movements as a network, in which edges form between outlets who share contributors. Using thousands of stories from 13 news outlets---with manually-validated information about the journalists, freelance writers, and political actors who wrote them---we demonstrate clear patterns in the trajectories contributors take when they write for multiple outlets. We then show that contributor trajectories across outlets align with the underlying political leanings of the outlets in our sample, more than any other characteristic of the outlets or their audiences. Furthermore, we show that those politically-aligned clusters differ in internal structure, with a highly interconnected cluster of right-leaning outlets. Finally, we link the political clustering present in this network to the content contributors produce. We find that, not only do both the topics writers cover in each cluster and the ways in which they present this content differ, but also that contributors that move between clusters tend to be less explicitly political. Taken together, these findings challenge the value of bridging content in reducing the polarization of media consumers. Rather, they demonstrate the need for a more nuanced examination of media \textit{producers} and their underlying motivations.

\section{Background}
\subsection{Polarization in News Consumption}
Digital news exists in a fragmented ecosystem. In contrast to mass media like broadcast or print, news consumers split their attention among a wide variety of sources \citep{websterMarketplaceAttention2016}. Audiences splinter into small groups as low-barrier, low-cost news sources tailor to their specific interests and identities \citep{tanejaPathwaysFragmentationUser2018}.

One prominent dimension of this fragmentation is political polarization. Past work has argued for the existence of ``filter bubbles'' or ``echo chambers'', completely isolated spheres of media consumption. The evidence for these kinds of all-encompassing divisions in news consumption is limited \citep{brunsAreFilterBubbles2019}. However, political beliefs and preferences are a clear \textit{factor} for consumers in news source and story selection. People generally prefer to read news that aligns with their political positions \citep{flaxmanFilterBubblesEcho2016}. Selective exposure along partisan lines also shapes readers' perceptions of news source credibility, thereby impacting what news they decide to engage with \citep{tsfatiExposureIdeologicalNews2014}.

Given this focus on \textit{audience} preferences and agency, polarization via fragmentation is treated primarily as a consumption behavior in the digital news ecosystem. The driving forces of polarization are often seen as news consumer choices, in terms of the sources that they select and the perspectives they prefer to see reflected in news coverage. In contrast, the countervailing forces to polarization are often treated as the purview of news \textit{producers}, who adhere to professional standards of balance and objectivity.

\subsection{Objectivity and Professionalism in News Production}
Many news organizations and individual journalists go to great lengths to avoid the appearance of any partisan position or influence. Historically, this was not the case. Up through the 19th century, U.S. newspapers explicitly identified themselves with political parties or ideologies, positions that influenced their coverage \citep{schudsonObjectivityNormAmerican2001}. 

As journalists developed a clearer and more cohesive professional identity throughout the 20th century, industry-wide norms developed that shifted news' presentation \citep{schudsonObjectivityNormAmerican2001}. In particular, journalists developed a strong link between professionalism and objectivity \citep{schudsonObjectivityProfessionalismTruth2009}. This objectivity seeks to balance the perspectives and positions represented in news coverage, rather than advancing any particular narrative \citep{ryanJournalisticEthicsObjectivity2001}. 

The practice of objectivity in journalism takes several forms at the individual and institutional level. For journalists, objectivity is often tied to professional independence---they value their ability to pursue stories without outside ideological influence \citep{beamChangesProfessionalismJournalists2009}. Striving for impartiality often homogenizes news, drawing coverage toward a political center that relies mostly on official sources to construct narratives \citep{mcchesneyProblemJournalismPolitical2003}. Objectivity is often an ethical ideal, not a value reflected in published news. This is particularly the case for dissenting or marginalized perspectives, which often get drowned out by these centralized, official narratives \citep{mcchesneyFAREWELLJOURNALISMTime2012}. Even so, objectivity, or at least its appearance, benefits news organizations as a whole. News outlets depend on credibility with their audience for sustainability \citep{singerStillGuardingGate1997}. Credibility and objectivity, while distinct concepts, are closely related: Credibility arises from an audience's perception of a news outlet's trustworthiness, and objectivity is a key measure of the amount of fairness or bias contained within coverage \citep{kellyEvaluatingNewsMis2019}. News organizations therefore implement policies and practices to appear more objective. The New York Times offers a productive case study. At the institutional level, it has marketed itself as an objective source of factual information, with slogans such as ``The truth is worth it''. At the individual level, it prohibits journalists from expressing ``partisan opinions'' or ``political views'' on social media \citep{thenewyorktimesTimesIssuesSocial2017}. These approaches complement each other, enforcing objectivity in practice and in the eyes of the audience. 

For institutional news sources and individual journalists, then, there is a strong incentive to uphold this notion of objectivity. News outlets require a level of political impartiality from their contributors, to avoid any appearance of bias. Journalists are encouraged to meet these standards by a professional code of ethics. And together, these norms and practices are marketed to audiences as safeguards for a credible, trustworthy source of news. Objectivity centers professional news production as the site of media de-polarization, by providing a common, reliable information source.

\subsection{Structural Polarization in News Distribution}
The work laid out so far tends to identify polarization as a news consumption trait, while construing objectivity and professionalism as traits of news production. This framing puts producers and consumers at odds, and it also suggests that de-polarization efforts should be aimed at audiences rather than creators. There has been much focus on ``bursting'' news readers' filter bubbles, by giving them indications of the partisan slant of their consumption habits or intentionally exposing them to opposing viewpoints \citep{resnickBurstingYourFilter2013,flaxmanFilterBubblesEcho2016}. This approach attempts to tackle partisan isolation in the digital news ecosystem by focusing on cross-cutting exposure in \textit{consumption}.

However, recent work has also examined the ways in which news \textit{production} and \textit{distribution}---particularly in the U.S.---becomes politically polarized. One driving aspect is commercial. Audiences engage most with perspectives that align with their own \citep{flaxmanFilterBubblesEcho2016}. In response, news outlets may gradually move further away from the center of the political spectrum, in an effort to better capture certain subsets of news readers \citep{mungerAllNewsThat2020,andersonMEDIAMERGERSMEDIA2012}. Algorithmic intermediaries also reinforce polarization via distribution, in that they reward more extreme offerings via recommendation \citep{ribeiroAuditingRadicalizationPathways2020, blex2020positive}.

Even beyond these mechanisms, researchers have identified \textit{structural} characteristics of the digital news ecosystem that reinforce polarization. In particular, \citet{benklerNetworkPropagandaManipulation2018} examine the pathways along which news stories travel among outlets. They find that, across several modes of distribution, right-leaning outlets tend to amplify each others' stories most often. They are also isolated from the rest of the media ecosystem, lacking ties, in terms of hyperlinks and social media sharing, to center- or left-leaning outlets. Outlets also diverge in the topics they cover, depending on their political leaning. In this way, \citet{benklerNetworkPropagandaManipulation2018} paint a picture of a media ecosystem that is polarized, not because of any individual agent or behavior, but because of a broad structure of distribution. This account runs counter to the idea of a professionalized, objective news media. It portrays an integrated system of polarization, one that cannot be addressed by focusing solely on the consumption patterns of news readers. It fundamentally addresses the \textit{portability} of news stories and perspectives among outlets and audiences, and how that is affected by polarization.

Our work takes a similar structural view of news production. Rather than examining the portability of news coverage, though, we are concerned with how polarization affects the movement patterns of news producers themselves. Given the strong individual focus on objectivity within the journalism profession, we are interested in whether the movement of individual writers among outlets is subject to the same structural pressures. This study addresses the following questions:

\noindent\textbf{RQ1:} What patterns do news contributors follow when moving among outlets?

\noindent\textbf{RQ2:} Do the topics addressed by contributors vary, depending on an outlet's political leaning?

\noindent\textbf{RQ3:} Do contributors who cross political boundaries cover different topics from those who do not?

\section{Data}
We started with a publicly-available data set of news articles scraped from the homepages of major publishers \citep{thompsonAllNews1432017}. The full data set contains 131,860 articles, published between June 2014 and July 2017, across 14 outlets: the New York Times, Breitbart, CNN, Business Insider, the Atlantic, Fox News, Talking Points Memo, Buzzfeed News, National Review, New York Post, the Guardian, NPR, Vox, and the Washington Post. We filtered this full data set in three ways. First, we removed any records with author fields that do not correspond to a person's name. To accomplish this, we removed any stories with bylines that matched generic phrases (e.g., ``the editors'', ``anonymous''), outlets (e.g., ``The Associated Press'', ``NPR staff''), or that contained bylines with multiple names. Second, because we are interested in movement between outlets, we limited our sample to articles by contributors who appeared in at least two outlets in the data set. Because Business Insider does not share any contributors with other outlets, we removed it from our analysis. Finally, we manually verified the remaining bylines, disambiguating cases of multiple people sharing the same name. This process identifies 368 contributors, who wrote 6,032 articles and are the focus of our analyses.

We next manually coded each contributor depending on their professional background. First, there are three types of professional journalists in our sample: Freelancers who write for multiple news outlets (97 names), journalists who are full-time employees at an outlet but write for at least one other (67 names), and journalists who move from one full-time job to another over the course of our sample (43 names). Overall, then, 207 of the contributors in our sample are professional journalists of some kind. We also find a couple groups who write pieces in support of particular issues or causes. This includes activists and think tank members (62 names) as well as political commentators and politicians (27 names). Finally, there are smaller groups of academics (33 names) and authors (33 names) who write pieces primarily to promote their work. An additional 6 contributors do not fit into these categories.

\begin{figure*}[!ht]
    \centering
    \begin{subfigure}[t]{0.32\textwidth}
        \centering
        \includegraphics[width=\linewidth]{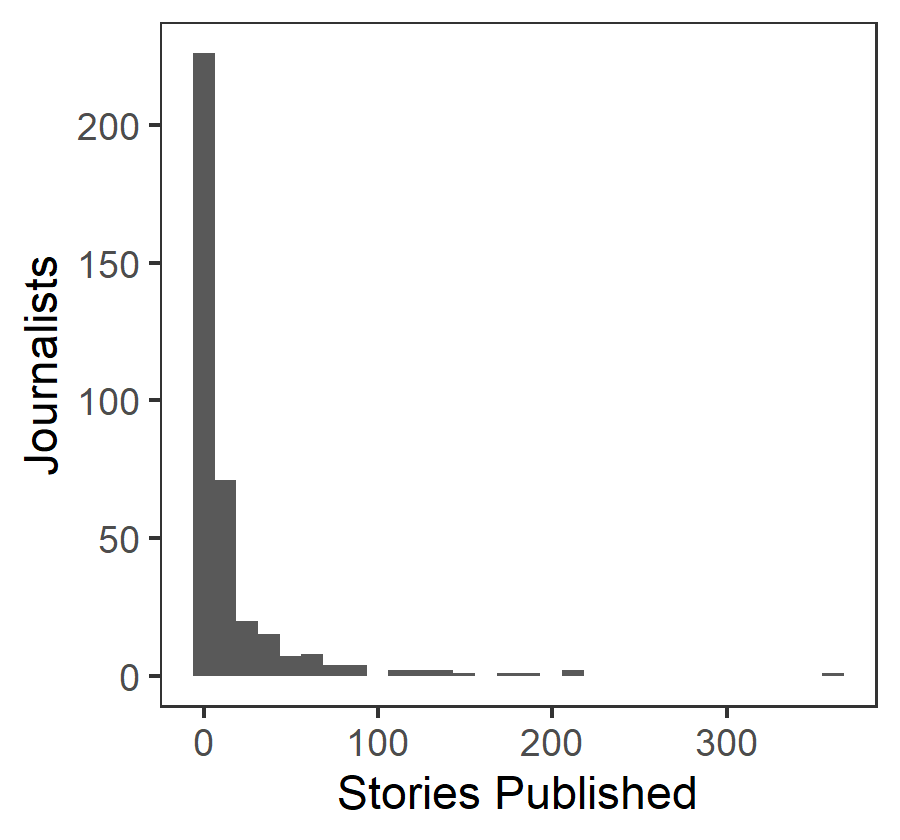}
        \caption{} \label{fig:stories}
    \end{subfigure}
    \hfill
    \begin{subfigure}[t]{0.32\textwidth}
        \centering
        \includegraphics[width=\linewidth]{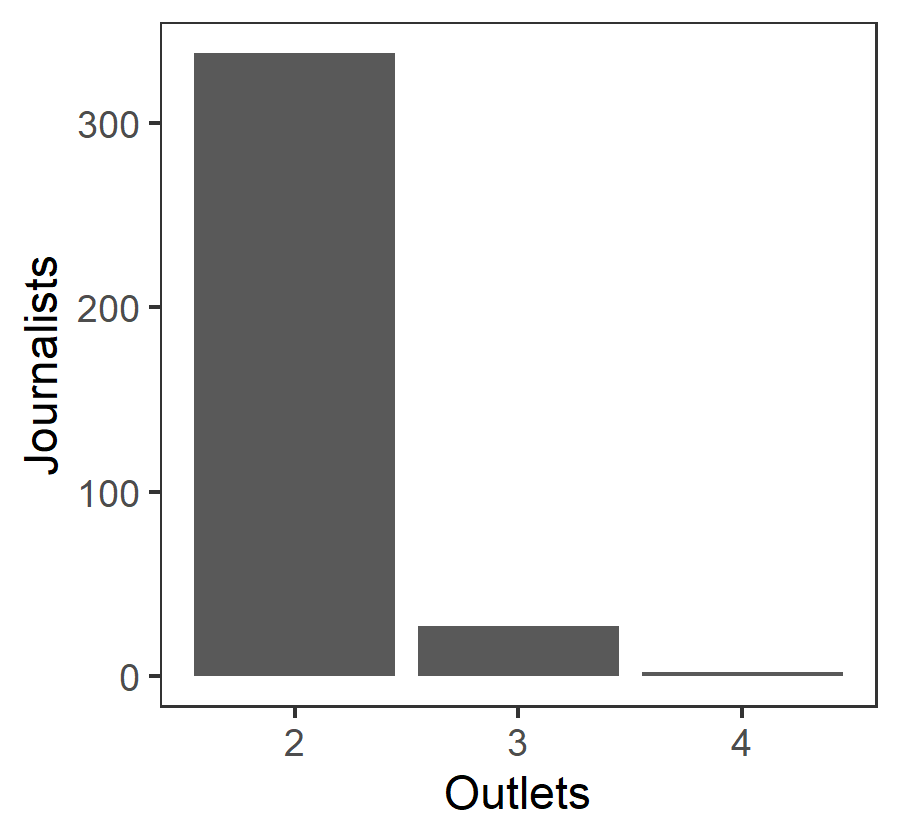}
        \caption{} \label{fig:pubs}
    \end{subfigure}
    \hfill
    \begin{subfigure}[t]{0.32\textwidth}
        \centering
        \includegraphics[width=\linewidth]{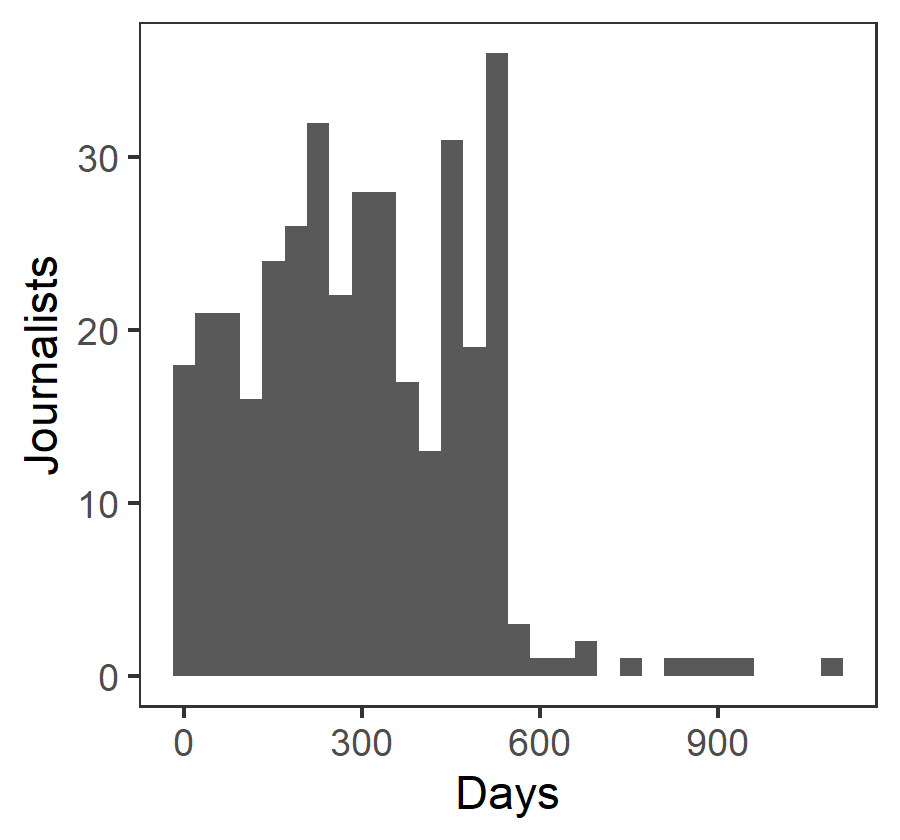}
        \caption{} \label{fig:days}
    \end{subfigure}
    \hfill
    \caption{Distributions of a) story counts per contributor, b) unique news outlets that published each contributor, and c) the number of days from each contributor's first to last story in our data.}
    \label{fig:descriptives}
\end{figure*}

Within our sample of 368 identified contributors, we analyzed publishing patterns for each individual (Figure \ref{fig}a). Figure \ref{fig:descriptives} presents the volume of stories, number of outlets published in, and number of days from first to last story for each contributor. We find that many contributors write across a couple outlets, and the majority publish fewer than 100 stories. However, there is widespread variation in how long contributors are actively publishing stories throughout our sample, indicating differences in their tempo of activity.

At the outlet level, we use a variety of characteristics to evaluate our eventual network's structure (Table \ref{classifications}). These characteristics fall under three broad types of measures, and they broadly capture information about the outlet and the size and composition of its audience. First, we used two measures of outlet size---total unique users and total time users spent reading each outlet---as these are often utilized in studies attempting to classify news outlets \citep[e.g.,][]{hindmanInternetTrapHow2018,mccombsDefiningLocalNews1981}. Second, we used a variety of audience characteristics---household size, whether or not a household has children, the race and gender composition of the audience, where the audience is located, and the median reader age and income. All measures except income report audience size, in terms of unique users, for certain groups. For example, household size reports total users for households with 1-2 and 3-5 members. All measures except region, age, and income report two groups, so we calculated the ratio of one group to the other to reduce the number of classifications. For audience region, we retained the largest region by number of users. Audiences often identify with the news they read, particularly in alternative media, which could create observable clusters of outlets within shared communities \citep{benklerNetworkPropagandaManipulation2018,kaiserConnectingFarRight2019,chiricosCrimeNewsFear1997,couldryContestingMediaPower2003}. Third we utilized a couple outlet characteristics: traditional publishing medium and political leaning. Publishing medium is a common classification approach for comparative work (e.g., \citep{maierAllNewsFit2010,boczkowskiWhenMoreMedia2007}), and while the articles we analyze are all published online, outlets still maintain other dominant distribution media that may inform their overall publishing strategy. We utilized the outlet political leanings generated by \citep{bakshyExposureIdeologicallyDiverse2015a} in examining polarization on Facebook.

ComScore provides all our audience measures. For these, we obtained the mean monthly value of each measure from January 2016 to December 2017, when most articles were collected. We then divided each continuous measure into three groups, each of the size $(max-min)/3$. For political leaning, we use a measure devised by \citet{bakshyExposureIdeologicallyDiverse2015a}. They analyze the self-reported political affiliation of Facebook users, assigning news articles an average alignment score based on who shared each article. They then average those article-level scores at the website level, producing a site-level alignment score ranging from -1 (left-leaning sharers) to 1 (right-leaning sharers). \citet{bakshyExposureIdeologicallyDiverse2015a} coarsen this measure by quintiles. We follow a similar approach, treating sites that fall outside of their central quintile as left- or right-leaning.

%\onecolumn
\begin{table}[!h]
\begin{tabular*}{0.65\linewidth}{lc}
\toprule
Measure & Source \\
\midrule
Total unique users & ComScore \\
Total time read & ComScore \\
Median reader income & ComScore \\
Median reader age & ComScore \\
Audience household size & ComScore \\
Audience children (yes/no ratio) & ComScore \\
Audience race & ComScore \\
Audience gender (M/F ratio) & ComScore \\
Audience region & ComScore \\
Outlet medium & Manual coding \\
Political leanings & See: \citep{bakshyExposureIdeologicallyDiverse2015a} \\
\bottomrule
\end{tabular*}
\caption{Measures used for classifying outlets, along with their sources.}
\label{classifications}
\end{table}

\section{Methods}
Here we describe the analytic approach we used to examine the network between outlets that maps news contributor movement, as well as to investigate the topic and tone of stories published.

\subsection{Contributor Network Structure}
To evaluate the network structure of contributor movement, we constructed a bipartite network of writers and outlets, utilizing only our filtered set of writers. A writer and a outlet shared an edge if at least one of the writer's stories appeared in the outlet within our sample. From this network, we constructed a one-mode projection for outlets, in which outlets share an edge if at least one writer published an article in both during the considered time frame (Figure \ref{fig}b). Edges were weighted according to the number of writers outlets share. The number of articles per contributor across outlets also varies; however, it does not vary evenly, and the article count is less relevant to our research question than the number of contributors shared. 31 pairs of outlets did not share any contributors.

We then compared this weighted projection to randomized bipartite networks of news outlets and contributors, to check for significant connections between outlets. We used a standard Markov Chain Monte Carlo algorithm to build random networks from the observed bipartite network \citep{cobbApplicationMarkovChain2003,tumminelloStatisticallyValidatedNetworks2011,newmanScientificCollaborationNetworks2001,raoMarkovChainMonte1996,zweigSystematicApproachOnemode2011}. We attempted $m*log_e(m)$ edge switches, where $m$ is the number of edges in the bipartite network (i.e., 784) to obtain one instance of a random network that has a substantially different structure than the observed network \citep{gionisAssessingDataMining2007}. We generated a set of 10,000 random networks. Our Supplementary Materials Table 1 show the significance of edges in the projection; we retain edges where $Z>1.96$. This procedure generates distinct clusters of outlets.

\subsection{Classification Evaluation}
To evaluate the uncovered network clustering, we compared it to classifications from the mixture of audience and outlet characteristics described in the Data section (Table \ref{classifications}). We treated each of these classifications as though they were partitions within our observed network, then used that partition to calculate the modularity of the classification. Ranging from -1 to 1, modularity describes the extent to which a network is split into distinct clusters \citep{newmanNetworksIntroduction2010,newmanModularityCommunityStructure2006}. By mapping classifications onto our observed network as though they were clusters, essentially we calculated the extent to which each one aligned with the observed structure. 

\subsection{Topic Modeling}
To analyze how the network structure of the journalism marketplace relates to content, we employed topic models~\citep{mohr2013introduction}. Specifically, we fit a topic model to the articles in our data set with Latent Dirichlet Allocation (LDA)~\citep{blei2002latent} using the Mallet program (and its default parameter settings)~\citep{mccallum2002mallet}, accessed via the gensim library~\citep{rehurek_lrec} of the Python programming language. The result assigned each document a vector of topics, allowing that a document consists of a combination of topics and that words appear in multiple topics. The number of topics is a parameter of the LDA algorithm--in the paper we presented results from a model set to find twenty topics. We found similar results, namely significant differences in topic prevalence across the clusters of outlets when we fit the model to 15, 25, and 30 topics.

Before we fit the topic model, we processed the text of the articles. We extracted each word or token, lower-casing all characters and removing apostrophes. We removed English-language stop words and stemmed the remaining words using the Porter stemmer. We removed words that occurred in more than one-third of articles and those that occurred in less than 10 articles in our corpus. The general goal of these steps was to reduce the noise of the data while keeping its main signals relevant to our research intact~\citep{hopkins2010method}. The order of our processing pipeline is in line with recent recommendations of best practices in topic modeling for communication research~\citep{maier2018applying}.

Our network analysis finds a clear two-cluster structure, one that maps onto outlets' political leanings. We checked whether there are significant differences in content between our derived outlet clusters by comparing the topic vectors of their articles. We identified which topics are over-represented among articles from individual clusters by calculating the average topic vector of articles in the cluster and then comparing this vector to average topic vectors calculated after randomizing the political lean of each article's outlet. This null model, which we generate 1,000 times, presents a random assignment of political label. For each topic, we calculated a Z-score comparing the prevalence of the topic in the observed articles with the average prevalence of the topic after randomization and scale by the standard deviation. We conducted this process separately for articles in each cluster.

\subsection{LIWC Features}
Besides the significant differences in content between the derived clusters, we also examine important stylistic differences. This inquiry is driven by the observed effect of news story presentation on audience reception, especially along partisan lines. In particular, past work has demonstrated how stories that are subjective and convey emotion reinforce partisan isolation in news consumption and increase virality on social media \citep{flaxmanFilterBubblesEcho2016,bergerWhatMakesOnline2012,xuWhatDrivesHyperPartisan2020}. Here we investigate how these semantic aspects of news stories vary between our outlet clusters. We apply the Linguistic Inquiry and Word Count (LIWC) dictionary \citep{tausczik2010psychological}, a widely used tool for investigating stylistic properties of text, to each article in our corpus. LIWC assigns text scores in various linguistic (e.g., the use of pronouns, prepositions, punctuation) and psychological (e.g., the use of words with significant positive or negative emotional valence) dimensions curated by human experts. LIWC features have been used to analyze the effectiveness of various kinds of persuasive writing from crowdfunding pitches \citep{horvat2018role} to public advocacy appeals \citep{bail2017channeling}. A study of democrats and republicans on Twitter using LIWC found significant differences between the characteristic linguistic style of the two groups \citep{sylwester2015twitter}, for instance that republicans were more likely to use words expressing negative emotion. 

Using the LIWC software, we scored each article in our corpus along 83 dimensions of linguistic features. We also considered two additional sentiment-based features from the VADER library \citep{hutto2014vader}. We compared the distribution of each feature between the politically-aligned clusters using a Mann-Whitney U test. We find that 50 out of 85 features have significant differences (Bonferroni corrected $p < .01$) in their distributions between the two groups of articles. The number of significant differences suggests substantial stylistic differences in the writing presented in the two clusters. The full table of features and differences are presented in Supplementary Materials Section 2.

\section{Results}
\begin{figure*}[ht]
    \centering
    \includegraphics[width=\linewidth]{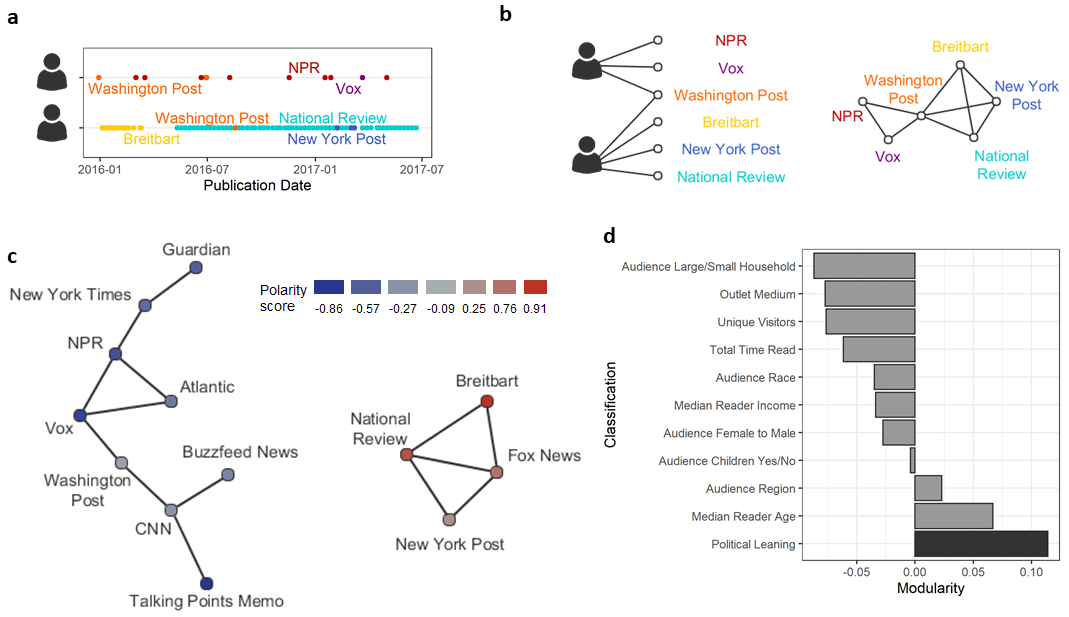}
    \caption{Mapping contributors' trajectories between news outlets. a) Example trajectories indicating the outlets for which the selected two contributors wrote articles. b) Illustration of how we build the network of outlets based on which pairs of outlets shared contributors. c) Significant connections between news outlets ($Z>1.96$), colored according to political leaning scores assigned by \citep{bakshyExposureIdeologicallyDiverse2015a}. d) Modularity for each outlet classification applied to the outlet network.}
    \label{fig}
\end{figure*}

To address RQ1, we first investigated the patterns common across contributors' movements. Since movement occurs at the level of the news outlet, we analyzed what contributors' next steps tended to be from any given outlet (e.g., do contributors who write a story for the \textit{New York Times} tend to also write for the \textit{Washington Post}?). Fig. \ref{fig}c demonstrates these trends, by showing our outlet-to-outlet network with significant edges only. This network of statistically significant movements between news outlets reveals striking patterns in contributors' publishing histories. In particular, two clusters of news outlets emerge: One low-density cluster comprised of nine outlets, and a dense cluster of four outlets. In the low-density cluster, only three outlets have connections with more than two outlets, creating a chain-like structure. From a contributor perspective, this indicates that jumps around the cluster are unlikely, but that travel to each outlet from every other is possible. In the high-density cluster, however, 5 of the 6 possible edges are present, indicating high mobility throughout the cluster. Between clusters, though, movement is highly unlikely. Outlets across clusters do not just lack significant positive connections; they have statistically significant negative edges. See the Supplementary Materials Table 1 for further details. 

This clustering structure suggests that there are some constraints or barriers to the movement of contributors between outlets, in particular as contributors seem to lack the flexibility to move between clusters. However, those constraints apply unevenly. Looking across contributor types, the majority of non-journalist authors, academics, and politicians cross between clusters at least once. In addition, politicians make up 36\% of all cross-cluster contributors, but only 5\% of single-cluster contributors. It seems that some kinds of news, particularly those associated with politics, travel easier than others.  However, less than a third of freelancers and other types of journalists move between clusters. For them, association with one cluster may represent a choice between two mutually-exclusive publishing paths, which have difficult-to-reverse consequences for future opportunities. Furthermore, these patterns are observable without taking into account any journalist or outlet attributes. Rather, they are structural characteristics that emerge from a combination of the editorial process and labor market. If the market structure of contributor-outlet relationships is built out of implicit patterns in activity, what heuristics---audience size or composition, for example---align with this structure? Are new classification mechanisms needed to understand the outlet landscape?

\subsection{Structure of Contributors' Movements between News Outlets}
Traditionally, work that examines multiple outlets has divided them by their publishing medium---comparing print newspapers to news websites, for example, or radio to television \citep{maierAllNewsFit2010}. In this network, though, the outlet's traditional medium does not seem to align with the clusters we observe: magazine contributors also write pieces for radio and digital-native sites, for example journalists shared by \emph{The Atlantic}, \emph{NPR}, and \emph{Vox}. To validate this observation, we categorized each outlet using a variety of metadata, described in the Data section. Then, we measured how well each of these classifications captures the clustering dynamic in our observed data using modularity (see details in Methods section). Our goal is to determine if traditionally-accepted classification strategies can capture similar information about contributor movement patterns as our network structure, and if not, whether any other characteristics might provide a useful proxy. 

In Figure \ref{fig}d, we show the modularity of each classification. Accordingly, outlet political leanings assigned by \citep{bakshyExposureIdeologicallyDiverse2015a} correspond best to the observed clusterings. If we map the leanings of each outlet to our observed network (Figure \ref{fig}c), we see a dense cluster of right-leaning outlets and a loose grouping of left- and center-leaning ones. Given this dichotomy, we subsequently refer to these clusters as left/center-leaning and right-leaning. This finding suggests that the political leaning of a news outlet is a key structural factor in where contributors will publish, pointing, as in prior work, to a polarized online news ecosystem \citep{benklerNetworkPropagandaManipulation2018,bakshyExposureIdeologicallyDiverse2015a,kaiserConnectingFarRight2019,wihbeySocialSilosJournalism2019,adamicPoliticalBlogosphere20042005}. While this structure impacts movement, though, does it affect the content of news articles across the network?

\subsection{News Coverage Effects}
The Z-scores derived from our topic model comparison across clusters (see Methods) indicate the most relevant words of topics statistically over-represented among left/center- and right-leaning outlets in the first columns of Table~\ref{table:topics}. A Z-score with absolute value greater than 1.96, corresponding to a p-value of 0.05, is considered a statistically significant deviation from the null hypothesis that cluster lean labels are unrelated to the distribution of topics. Articles appearing in center/left-leaning outlets are significantly more likely to be about science and research (Topic 8), the media and press (Topic 3), and healthcare (Topic 16), among others. In right-leaning outlets, articles are more likely to be about conservatism and liberalism (Topic 10), prominent Democratic Party politicians (Topic 13), and Republican politics (Topic 4).

17 out of 20 topics are significantly over-represented among writings appearing in either the right or left-leaning clusters of outlets. In relation to RQ2, this indicates that the partition of outlets according to transitions by contributors aligns with differences in content. Our findings prompt a question: Do those contributors that do transition between the left and right-leaning clusters adapt their writing topics to the venue? Or do these transitioning contributors fill in particular niches within the partisan spheres they visit?

To address this question, consider three groups of contributors: those who write only for center/left-leaning outlets, those who write only for right-leaning outlets, and those who write for both. Prior literature suggests that these groups will focus on different topics, for two reasons. First, if outlets gravitate toward certain political positions, they will often emphasize pieces of information relevant and favorable to that political group, and omit those that are not \citep{bernhardtPoliticalPolarizationElectoral2008}. Therefore, if contributors write stories that are in line with the editorial approach of the outlet as a whole, we would also expect the information they produce to shift depending on their group of focus. 

Second, it is challenging for either individual journalists or outlets as institutions to credibly maintain multiple political positions in their work \citep{andersonMEDIAMERGERSMEDIA2012}. A contributor who writes a piece about the dangers of climate change, for example, cannot also credibly write a piece denying its existence. It is then in contributors' best interest to establish and develop a consistent perspective in their writing throughout their careers. The contributors who remain in only one cluster will generate that cluster's representative content (e.g., freelancers publishing in left-leaning outlets will focus on the media), while contributors who move between clusters will write about topics that aren't particular to either. In other words, in response to RQ3, we hypothesize that contributors moving between the left and right-leaning outlets tend to write about politically more neutral topics.

\begin{landscape}
\begin{table*}[t]
\begin{xtabular*}{\linewidth}{lrrrr}
\toprule%
 \multicolumn{1}{l}{{{\bfseries Topic: Keywords}}}
 & \multicolumn{1}{c}{{{\bfseries CntrLeft Z}}}
 & \multicolumn{1}{c}{{{\bfseries Right Z}}}
 & \multicolumn{1}{c}{{{\bfseries Trans-CntrLeft Z}}}
 & \multicolumn{1}{c}{{{\bfseries Trans-Right Z}}}
 \\
\cmidrule[0.3pt](r{0.3em}){1-1}%
\cmidrule[0.3pt](lr{0.3em}){2-2}%
\cmidrule[0.3pt](lr{0.3em}){3-3}%
\cmidrule[0.3pt](lr{0.3em}){4-4}%
\cmidrule[0.3pt](lr{0.3em}){5-5}%}

1: attack, isi*, islam, muslim, war  & 0.42 & 0.8 &  \textit{-3.69}& \textit{-5.29} \\
\myrowcolour%
2: vote, voter, percent, parti*, poll & \textit{-2.63} & \textbf{4.07} &0.52 & 0.58  \\
3: news, media, post, press, twitter & \textbf{9.99} & \textit{-9.97} & \textit{-5.17} & -1.16 \\
\myrowcolour%
4: senat*, cruz, candid*, gop, parti*& \textit{-5.62} & \textbf{7.24} & \textbf{2.20} & \textbf{4.70}\\
5: realli*,talk, didn*, lot, someth* & \textbf{4.75} & -0.86 & \textit{-5.55}& 0.38 \\
\myrowcolour%
6: game, play, team, season, serv* & \textbf{6.31} & \textit{-5.65} &  1.86 & \textbf{6.85}\\
7: women, school, children, famili*, student& -1.04 & \textbf{3.16} & \textit{-4.33} & \textbf{5.62} \\
\myrowcolour%
8: research, studi*, human, found, univers* & \textbf{13.07} & \textit{-12.37} & \textit{-7.51} & \textbf{2.18} \\
9:  film, music, play, movi*, book& \textbf{5.23} & \textit{-3.47}&\textit{-2.66}& \textbf{3.37} \\
\myrowcolour%
10: conserv*, power, left, liber*,fact& \textit{-23.59} & \textbf{27.43} & \textbf{11.93}& \textit{-4.57}  \\
11: compani*, million, bank, busi*, billion& \textit{-6.64} & \textit{-8.7} & \textit{-3.24}&\textbf{13.07}  \\
\myrowcolour%
12: court, law, immigr*, case, rule& -0.36 & -0.76 & \textbf{18.86}& -1.25 \\
13: clinton, obama, hillari*, bill, email &\textit{-15.61} & \textbf{14.45} & \textbf{2.3} & \textit{-6.63} \\
\myrowcolour%
14: white, black, christian, commun*, religi* & \textit{-3.3} & \textbf{5.38} & -1.29 & \textit{-2.86} \\
15: tax, percent, job, econom*, rate & \textit{-3.92} &  \textbf{3.16} & -0.28 & -0.91 \\
\myrowcolour%
16: health, care, plan, bill, insur* & \textbf{7.7} & \textit{-7.14} & \textit{-3.64} & -0.02 \\
17: china, obama, unit, iran, foreign & \textbf{3.11} &  \textit{-2.89} & \textbf{6.92} & 1.35  \\
\myrowcolour%
18: offici*, investig*, secur*, inform, depart &\textbf{6.2} & \textit{-6.21} & \textit{-2.68} & \textit{-5.58} \\
19: citi*, build, water, climat*,area& \textbf{4.69} &  \textit{-5.1} & -1.54 & -0.65  \\
\myrowcolour%
20:  polic*, offic*, crime, gun, case & \textit{-2.06} & \textbf{2.03} & -0.50 & \textit{-4.51} \\
\bottomrule\\
\end{xtabular*}
\caption{Topic keywords and their statistical over-representation (bold) and under-representation (italic) within news outlet clusters. The first two columns of Z-scores present statistical over/under-representation of topics within the center/left- and right-leaning clusters. The last two columns record the statistical over/under-representations of topics in articles by contributors who move between clusters, compared to those by contributors who stay within the center/left- and right-leaning clusters, respectively.}
\label{table:topics}
\end{table*}
\end{landscape}

We tested this idea by considering within-cluster differences. For both the left and right-leaning cluster of outlets we compared two kinds of contributors: those who stayed within the cluster, which we call purists, and those who transitioned. For example, among all articles in right-leaning outlets we compare which topics tend to be covered by purists and those which tend to be covered by transitioning contributors. We repeated a similar experiment to the one used to determine topics over-represented among left and right-leaning outlets. Within a cluster we calculated the average topic vector of articles written by purists. We compared the average topic vector to those calculated from 1,000 randomizations realized by shuffling the purist/transitioning contributor labels. This randomization allowed us to test the statistical over-representation of topics by purists or transitioning contributors within the partisan clusters.

The results of this investigation are shown in the last two columns of Table~\ref{table:topics} for articles in left-leaning outlets and right-leaning outlets, respectively. Here we report the relative prevalence of topics in articles written by transitioning authors within the two clusters, compared to their purist counterparts. In both clusters there are many topics which are significantly over and under-represented in the writings of transitioning authors. This indicates that switching contributors in both partisan clusters occupy specific niches with their writing. 

Among the articles in right-leaning outlets, transitioning contributors are more likely to write about finance (Topic 11), sports (Topic 6), and families (Topic 7), among other topics. They are significantly less likely to write about Democratic Party politicians (Topic 13), investigations (Topic 18), and conflict in the Middle East (Topic 1). We note that of the six topics significantly under-represented in the writings of transitioning contributors relative to purist, three topics are globally over-represented in right-leaning outlets (Topics 10, 13, 14). We interpret this as suggesting that transitioning contributors are less likely to write about extremely partisan topics.

The same analysis of left-leaning outlets suggests a similar pattern. Transitioning contributors writing for left-leaning outlets seem to avoid topics over-represented in those outlets on a global level (Topics 3, 8, 18). Two of the five topics in which switching contributors are over-represented are globally over-represented among right-leaning outlets. Future work should investigate whether this observed relationship is because transitioning contributors avoid partisan topics in order to remain widely employable, or perhaps because the kind of contributor who is able to switch between the partisan clusters tends to specialize in these topics.

\subsection{Stylistic Differences in Language}
Our LIWC and sentiment analysis also generate clear differences across outlet clusters. We report the top six most distinguishing features by effect size (measured by the AUC) in Table~\ref{table:bigeffectliwc}; the full table of features can be found in our Supplementary Materials, Section 2. Three of these features are significantly over-represented in articles in center/left-leaning outlets: \textit{hear}, capturing the use of words describing the act of listening; \textit{percept}, capturing the use of observational words relating to perception; and \textit{focuspast}, capturing the use of past-tense verbs. Three are over-represented in right-leaning outlets: \textit{affect}, counting the use of words with significant psychological content; \textit{negemo}, counting the frequency of words with negative emotional connotation; and \textit{certain}, counting the use absolute words such as ``always'' or ``never''.

%\onecolumn
\begin{table*}
\tablehead{\toprule
Feature &        M-W U  &       AUC &  CntrLeft Avg. &  Right Avg. &  Diff. &  Bonf. P \\
\midrule}
\bottomcaption{The top six LIWC features distinguishing texts from center/left- (bold) and right-leaning (italic) outlets, by their AUC scores.}
\label{table:bigeffectliwc}
\begin{xtabular*}{0.85\linewidth}{lccccccc}
     \textbf{hear} &  3144691.0 &   0.65 &          1.08 &           0.71 &  0.37 &       $< 10^{-89}$      \\
    \textit{affect} &  3242798.0 &  0.64 &          4.33 &           5.10 & -0.77 &        $< 10^{-77}$     \\
    \textbf{percept} &  3260857.5 &   0.64 &          2.38 &           1.87 &  0.51 &        $< 10^{-75}$ \\
   \textit{negemo} &  3438952.5 &  0.62 &          1.88 &           2.36 & -0.48 &         $< 10^{-55}$  \\
    \textit{certain} &  3476109.5 &   0.62 &          1.06 &           1.30 & -0.24 &        $< 10^{-51}$     \\
  \textbf{focuspast} &  3616885.0 &    0.60 &          4.09 &           3.45 &  0.64 &            $< 10^{-38}$ \\
\bottomrule
\end{xtabular*}
\end{table*}
%\twocolumn

At a high level, then, we observe widespread differences in the relative usage of certain linguistic and semantic markers across right- and center/left-leaning outlets, giving us an affirmative answer to RQ2. Not only are audiences across these clusters exposed to different \textit{content}; that content is also subject to differing \textit{presentations}. In particular, we see that the usage of affect and emotion varies across these clusters, suggesting that the subjective frames with which writers present news stories depend on the preferences of the outlet and its audience.

\section{Discussion}
This work addresses an understudied area of news media polarization: structural production forces driving partisan leanings. By constructing a cross-outlet network purely based on contributor movement patterns, we show a clear partisan divide within the digital news ecosystem. That divide appears without any explicit consideration of audience behavior or preferences. 

Our results demonstrate the ways in which structural factors can work against individual ones, such as a professional commitment to objectivity. Some contributors in our sample are not journalists, and some are explicitly political figures. However, those that do fall under the umbrella of professional journalism often stay within partisan bounds. Somewhere within the editorial process of pitching, selecting contributors, assigning stories, and producing news coverage, a dynamic arises that structurally prefers contributors whose publishing histories ideologically align with a publication's own. This may arise from institutional policies, from individual editors' preferences, or from the pitching process of individual contributors \citep{christinMakingPeaceMetrics2020,rosenkranzContractSpeculationNew2018}. In fact, the driving mechanisms may differ between outlets or individuals. These possible mechanisms warrant further study. Given the ways in which objectivity is so forcefully conveyed as a key standard at many outlets, it is also worth examining where outlet interactions with structural forces causes this norm to break down.

Polarized contributor isolation also shows up in the topic and style of news coverage they produce, wherein distinct areas of interest and modes of presentation arise for each outlet cluster. The polarization effect might originate with contributor movement, but it also affects news coverage production at the story level. Prior work has demonstrated differences in news coverage approaches across the political spectrum \citep{xuWhatDrivesHyperPartisan2020,schifferAssessingPartisanBias2006}. But the key here is that those differences don't exist in isolation, but are rather part of a larger media system that reinforces itself to drive polarization. Because of the integrated nature of polarization within this system, we cannot say what ``causes'' it. We can point to specific locations or mechanisms by which it appears, such as selective exposure or profit maximization at the outlet level \citep{stroudPolarizationPartisanSelective2010,andersonMEDIAMERGERSMEDIA2012}. However, saying any of those things \emph{cause} polarization, or are even the main site of polarization, runs the risk of leading to pinpoint interventions that are broadly ineffectual. Audience-targeted interventions ignore producers' incentives to publish traffic-driving partisan coverage. Producer-targeted interventions ignore audiences' desires for similar perspectives to their own. Neither fully grapples with the feedback mechanism—running through the production, distribution, and consumption of digital news--tied to revenue incentives outlets must acknowledge to remain sustainable \citep{andersonMEDIAMERGERSMEDIA2012, mcchesneyFAREWELLJOURNALISMTime2012}. Future work should aim to work toward holistic interventions that address this integrated polarization.

We also identify topics less related to politics that are more common for contributors who move between clusters, a potential area for exploration in future work. In particular, it would be worthwhile to investigate whether the commonality of some topics also holds from an audience perspective--does sports coverage cut across political preferences, or does the partisan affiliation of a particular outlet affect all the coverage it produces?

Finally, in addition to addressing the broader area of news polarization research, this study also engages directly with structural production polarization work. \citet{benklerNetworkPropagandaManipulation2018} identify a very similar network structure in news media distribution to what we see here: a loosely connected group of left/center-leaning outlets, a dense core of right-leaning outlets, and very little activity between. Using a totally separate sample, and examining a different aspect of news production, we find an ecosystem with identical characteristics. \citet{benklerNetworkPropagandaManipulation2018} also demonstrate how the periphery of right-leaning outlets amplify conspiracy theories and misinformation into the core of mainstream media. In contrast, we see very little movement from this periphery outward, as contributors are significantly unlikely to move between clusters. This network structure is potentially good for stemming the flow of misinformation, but potentially bad for amplifying the content-based partisan leanings we see in both clusters.

\subsection{Limitations}
This study has several important limitations. First, we cannot control for all the potential biases that may have arisen in data collection via scraping. We demonstrate the robustness of our results to differences in sample and significance level with additional checks. First, we check the robustness of the network to varying Z-score thresholds (Supplementary Materials, Section 1.1). We also examine the impact of removing individual outlets from the network on its structure (Supplementary Materials, Section 1.2). However, it is still important to acknowledge potential biases introduced by this dataset.

Similarly, our sample is only a subset of all news outlets. While the outlets used here are prominent sources of digital media from across the political spectrum, they are only one part of the news ecosystem. Future work should attempt to broaden the outlets used.

Finally, outlet political leaning is a difficult quality to measure. It originates from the perspective of the audience, rather than any objective measure of the news coverage an outlet produces. We use a rigorous, peer-reviewed method to ascertain our scores \citep{bakshyExposureIdeologicallyDiverse2015a}, but other approaches may categorize outlets differently.

\subsection{Conclusions}
This study demonstrates the polarized structure of digital news contributor movement. It ties that structure to differences in news coverage content and perspective across the political spectrum, and it highlights the topic differential across inter- and intra-cluster movement. Together, these results show the interconnected nature of polarization and its consequences across various aspects of news production. These findings fill an important gap in the literature on news media, highlighting structural factors on the production side which may contribute to polarization.

\bibliographystyle{agsm}
\bibliography{Journal-bib-freelancers.bib}
\section{Supplementary Materials}

\subsection{Network robustness}
\subsubsection{Z-score threshold}
We recognize that there are potential sensitivities in our analysis to the Z-score threshold we have chosen. In particular, it is possible that edges were barely excluded at the $Z>1.96$ threshold that would change some of the clustering dynamic we observe, or that edges crucial to the structure of the clusters would disappear at a stricter threshold. To account for these scenarios, we construct additional networks with edges selected by varying Z-scores. In addition to the network we present in the main text, which filters out edges based on a $Z>1.96$ threshold, we also check 1.64, 2.58, and 3.29. No additional analyses break down the clustering dynamic we observe in our initial network. We present the underlying values for each network edge in Table \ref{table:raw_results}.

At $Z>1.64$, one edge gets added to the left-leaning cluster (between \textit{The Guardian} and \textit{Vox}). Neither the clusters themselves or the underlying structures--a dense right-leaning cluster versus a more chain-like left-leaning cluster--are disturbed.

At $Z>2.58$, eight of the 24 significant edges in our initial projection are lost. Four of these are negative edges across clusters, and three are in the left-leaning cluster. This means that the structure of the left leaning cluster becomes weaker, but, while the two clusters no longer have a significant negative association, they are still not connected. 

At $Z>3.29$, eight additional edges are lost. Once again, we see the weakening of the left-leaning cluster, as well as the loss of negative edges between clusters. Strikingly, the right-leaning cluster remains almost entirely intact, even at this level.

\newpage
\bottomcaption{Underlying frequencies of shared contributors between news outlets and p-values/Z-scores generated by SICOP}
\label{table:raw_results}
\tablehead{\toprule
From & To & Shared journalists & p & Z-score \\
\midrule}
\begin{xtabular*}{\linewidth}{llrrr}
    NationalReview & NewYorkPost & 43 & 0.00 & 9.26 \\ 
    NationalReview & WashingtonPost & 24 & 0.86 & -0.18 \\ 
    NationalReview & Atlantic & 6 & 0.01 & -2.50 \\ 
    NationalReview & BuzzfeedNews & 1 & 0.25 & -1.14 \\
    NationalReview & Breitbart & 18 & 0.00 & 7.41 \\ 
    NationalReview & FoxNews & 13 & 0.00 & 3.52 \\ 
    NationalReview & Guardian & 2 & 0.00 & -3.07 \\ 
    NationalReview & Vox & 1 & 0.00 & -2.88 \\ 
    NewYorkPost & NewYorkTimes & 2 & 0.01 & -2.57 \\ 
    NewYorkPost & WashingtonPost & 16 & 0.04 & -2.08 \\
    NewYorkPost & Atlantic & 2 & 0.00 & -3.60 \\ 
    NewYorkPost & BuzzfeedNews & 1 & 0.30 & -1.03 \\ 
    NewYorkPost & Breitbart & 8 & 0.03 & 2.24 \\ 
    NewYorkPost & FoxNews & 22 & 0.00 & 8.23 \\ 
    NewYorkPost & Guardian & 4 & 0.01 & -2.48 \\ 
    NewYorkPost & NPR & 1 & 0.00 & -3.20 \\ 
    NewYorkTimes & WashingtonPost & 17 & 0.54 & 0.61 \\
    NewYorkTimes & Atlantic & 10 & 0.57 & 0.57 \\ 
    NewYorkTimes & BuzzfeedNews & 2 & 0.78 & 0.28 \\ 
    NewYorkTimes & Guardian & 16 & 0.00 & 4.15 \\ 
    NewYorkTimes & NPR & 13 & 0.00 & 2.89 \\ 
    NewYorkTimes & Vox & 4 & 0.56 & -0.58 \\ 
    WashingtonPost & Atlantic & 28 & 0.38 & 0.88 \\ 
    WashingtonPost & BuzzfeedNews & 3 & 0.44 & -0.78 \\
    WashingtonPost & Breitbart & 5 & 0.18 & -1.33 \\ 
    WashingtonPost & FoxNews & 2 & 0.01 & -2.79 \\ 
    WashingtonPost & Guardian & 17 & 0.97 & -0.04 \\ 
    WashingtonPost & NPR & 20 & 0.37 & 0.89 \\ 
    WashingtonPost & Vox & 24 & 0.00 & 3.08 \\ 
    WashingtonPost & CNN & 8 & 0.01 & 2.45 \\ 
    WashingtonPost & TalkingPointsMemo & 1 & 0.34 & 0.95 \\ 
    Atlantic & BuzzfeedNews & 3 & 0.74 & 0.33 \\ 
    Atlantic & Guardian & 15 & 0.05 & 1.97 \\ 
    Atlantic & NPR & 21 & 0.00 & 4.16 \\ 
    Atlantic & Vox & 16 & 0.00 & 3.28 \\ 
    Atlantic & CNN & 3 & 0.35 & 0.94 \\ 
    BuzzfeedNews & Guardian & 4 & 0.10 & 1.64 \\ 
    BuzzfeedNews & NPR & 1 & 0.49 & -0.69 \\ 
    BuzzfeedNews & Vox & 1 & 0.67 & -0.43 \\ 
    BuzzfeedNews & CNN & 2 & 0.01 & 2.77 \\ 
    Breitbart & FoxNews & 6 & 0.00 & 3.67 \\ 
    Guardian & NPR & 8 & 0.81 & 0.24 \\ 
    Guardian & Vox & 10 & 0.06 & 1.90 \\ 
    NPR & Vox & 11 & 0.02 & 2.32 \\ 
    NPR & CNN & 1 & 0.73 & -0.35 \\ 
    Vox & CNN & 1 & 0.93 & -0.09 \\ 
    CNN & TalkingPointsMemo & 1 & 0.00 & 4.98 \\ 
\bottomrule
\end{xtabular*}
%\end{table*}

\newpage

\subsection{Outlets included}
Because our data are collected secondhand, they are subject to a couple potential biases that might affect our results. First, the selection of outlets within the sample may shape the network structure. Second, unobserved inconsistencies in data collection across outlets could distort the contributor publishing histories we use to evaluate edge significance. To better understand the potential impact of these factors, we removed one outlet at a time from our sample, then ran SICOP again on each set of 12 remaining outlets. Rather than just removing a node from our final network, this procedure allows new edge weights to be calculated without the influence of a particular outlet.

\begin{figure*}[ht]
    \centering
    \begin{subfigure}[t]{0.48\textwidth}
        \centering
        \includegraphics[width=\linewidth]{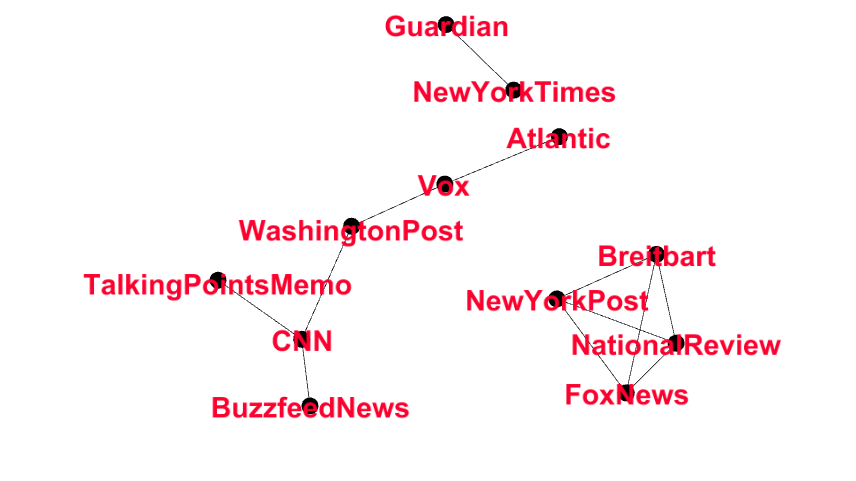}
        \caption{} \label{fig:stories}
    \end{subfigure}
    \hfill
    \begin{subfigure}[t]{0.48\textwidth}
        \centering
        \includegraphics[width=\linewidth]{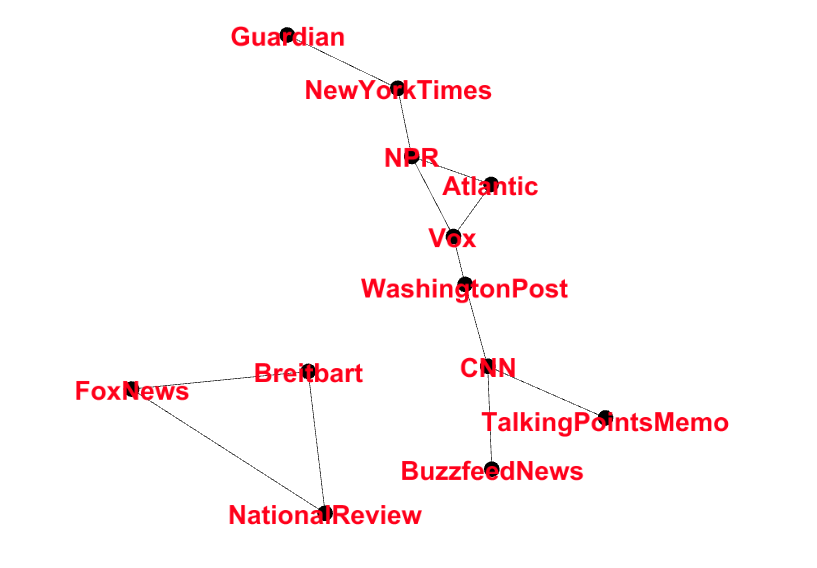}
        \caption{} \label{fig:pubs}
    \end{subfigure}
    \hfill
    \caption{Outlet-outlet projections with a) NPR removed and b) the New York Post removed. These cases are representative of the minor changes in network structure created by dropping outlets from our sample, maintaining the characteristics we highlight in our findings.}
    \label{fig:removals}
\end{figure*}

In no case did the overall network structure we observe drastically change as a result of this procedure. We still see a loose collection of left-/center-leaning outlets, and a dense cluster of right-leaning ones. In some cases, the left/center cluster breaks into multiple clusters, or into one cluster with isolates (e.g., Fig. \ref{fig:removals} a). This follows naturally from the cluster's observed chain-like structure---removing central nodes causes the chain to break. However, none of these iterations change the looseness of the overall structure. Similarly, in the right-leaning cluster, removing any one of the four nodes simply causes edges to form among the other three (e.g., Fig. \ref{fig:removals}b). Most importantly, in no iteration do any significant edges form between clusters. Thus, while the particular structure of the left/center cluster does show some sensitivity to the outlets included, the overall division and the characteristics of the clusters within this network remain consistent.

\newpage
\subsection{LIWC}

Here we report the full list of LIWC (and VADER) features compared between the two primary groups of articles. The entries are sorted by AUC.

\begin{longtable}{lrrrrrrr}
\toprule
Feature &        M-W U &          P-val. &       AUC &  CntrLeft Avg. &  Right Avg. &  Diff. &  Bonf. P-val. \\
\midrule

          hear &  3144691.0 &  0.000 &  0.652 &          1.08 &           0.71 &  0.37 &                   0.000 \\
        affect &  3242798.0 &  0.000 &  0.641 &          4.33 &           5.10 & -0.77 &                   0.000 \\
       percept &  3260857.5 &  0.000 &  0.639 &          2.38 &           1.87 &  0.51 &                   0.000 \\
        negemo &  3438952.5 &  0.000 &  0.619 &          1.88 &           2.36 & -0.48 &                   0.000 \\
       certain &  3476109.5 &  0.000 &  0.615 &          1.06 &           1.30 & -0.24 &                   0.000 \\
     focuspast &  3616885.0 &  0.000 &  0.600 &          4.09 &           3.45 &  0.64 &                   0.000 \\
         anger &  3625115.0 &  0.000 &  0.599 &          0.69 &           0.91 & -0.22 &                   0.000 \\
       relativ &  3630524.5 &  0.000 &  0.598 &         13.63 &          12.72 &  0.91 &                   0.000 \\
        negate &  3656126.5 &  0.000 &  0.595 &          1.16 &           1.36 & -0.20 &                   0.000 \\
        posemo &  3780913.0 &  0.000 &  0.582 &          2.37 &           2.67 & -0.30 &                   0.000 \\
         QMark &  3797283.5 &  0.000 &  0.580 &          0.18 &           0.25 & -0.07 &                   0.000 \\
         space &  3797893.0 &  0.000 &  0.580 &          7.28 &           6.82 &  0.46 &                   0.000 \\
           sad &  3818021.5 &  0.000 &  0.577 &          0.29 &           0.37 & -0.08 &                   0.000 \\
       discrep &  3839167.0 &  0.000 &  0.575 &          1.19 &           1.37 & -0.18 &                   0.000 \\
        differ &  3877426.5 &  0.000 &  0.571 &          2.73 &           2.97 & -0.24 &                   0.000 \\
         Comma &  3878505.0 &  0.000 &  0.571 &          5.73 &           5.39 &  0.34 &                   0.000 \\
          risk &  3897427.0 &  0.000 &  0.569 &          0.71 &           0.82 & -0.11 &                   0.000 \\
          prep &  3902945.5 &  0.000 &  0.568 &         14.57 &          14.18 &  0.39 &                   0.000 \\
          conj &  3913954.5 &  0.000 &  0.567 &          5.31 &           5.54 & -0.23 &                   0.000 \\
 vaderCompound &  3928249.5 &  0.000 &  0.565 &          0.25 &           0.11 &  0.14 &                   0.000 \\
         power &  3936100.0 &  0.000 &  0.564 &          4.23 &           4.55 & -0.32 &                   0.000 \\
       cogproc &  3950743.5 &  0.000 &  0.563 &          9.58 &          10.09 & -0.51 &                   0.000 \\
          time &  3982288.5 &  0.000 &  0.559 &          4.69 &           4.38 &  0.31 &                   0.000 \\
        motion &  3997062.0 &  0.000 &  0.558 &          1.74 &           1.60 &  0.14 &                   0.000 \\
       auxverb &  4016272.5 &  0.000 &  0.555 &          6.77 &           7.05 & -0.28 &                   0.000 \\
           see &  4063008.0 &  0.000 &  0.550 &          0.87 &           0.77 &  0.10 &                   0.000 \\
         Quote &  4110413.5 &  0.000 &  0.545 &          2.48 &           2.28 &  0.20 &                   0.000 \\
        drives &  4116632.5 &  0.000 &  0.544 &          8.46 &           8.79 & -0.33 &                   0.000 \\
  focuspresent &  4132848.0 &  0.000 &  0.543 &          7.19 &           7.46 & -0.27 &                   0.000 \\
        assent &  4145328.5 &  0.000 &  0.541 &          0.06 &           0.09 & -0.03 &                   0.000 \\
           adj &  4152054.5 &  0.000 &  0.540 &          4.77 &           4.94 & -0.17 &                   0.000 \\
         death &  4158694.5 &  0.000 &  0.540 &          0.24 &           0.27 & -0.03 &                   0.000 \\
        social &  4164598.0 &  0.000 &  0.539 &          8.96 &           8.55 &  0.41 &                   0.000 \\
         relig &  4164692.0 &  0.000 &  0.539 &          0.33 &           0.41 & -0.08 &                   0.000 \\
             i &  4164564.0 &  0.000 &  0.539 &          0.71 &           0.53 &  0.18 &                   0.000 \\
    tbPolarity &  4178439.5 &  0.000 &  0.538 &          0.09 &           0.08 &  0.01 &                   0.000 \\
        reward &  4181480.0 &  0.000 &  0.537 &          1.00 &           1.07 & -0.07 &                   0.000 \\
          work &  4188183.5 &  0.000 &  0.536 &          4.60 &           4.31 &  0.29 &                   0.000 \\
          home &  4195092.5 &  0.000 &  0.536 &          0.40 &           0.33 &  0.07 &                   0.000 \\
       Apostro &  4190130.0 &  0.000 &  0.536 &          2.22 &           2.31 & -0.09 &                   0.000 \\
         SemiC &  4208843.5 &  0.000 &  0.534 &          0.03 &           0.02 &  0.01 &                   0.000 \\
          they &  4219078.0 &  0.000 &  0.533 &          0.86 &           0.95 & -0.09 &                   0.000 \\
         money &  4232751.5 &  0.000 &  0.532 &          1.12 &           1.23 & -0.11 &                   0.001 \\
         quant &  4235906.0 &  0.000 &  0.531 &          2.16 &           2.24 & -0.08 &                   0.001 \\
        adverb &  4241383.5 &  0.000 &  0.531 &          3.54 &           3.64 & -0.10 &                   0.002 \\
        ingest &  4234582.0 &  0.000 &  0.531 &          0.26 &           0.19 &  0.07 &                   0.000 \\
      netspeak &  4249819.0 &  0.000 &  0.530 &          0.21 &           0.18 &  0.03 &                   0.001 \\
      interrog &  4244529.0 &  0.000 &  0.530 &          1.44 &           1.39 &  0.05 &                   0.002 \\
       AllPunc &  4255325.5 &  0.000 &  0.529 &         17.26 &          16.84 &  0.42 &                   0.004 \\
        OtherP &  4258541.5 &  0.000 &  0.529 &          0.28 &           0.30 & -0.02 &                   0.002 \\
       leisure &  4274775.5 &  0.000 &  0.527 &          1.03 &           0.90 &  0.13 &                   0.013 \\
      function &  4314285.5 &  0.001 &  0.523 &         46.19 &          46.21 & -0.02 &                   0.110 \\
        tentat &  4330995.5 &  0.003 &  0.521 &          2.18 &           2.25 & -0.07 &                   0.241 \\
           anx &  4336993.5 &  0.004 &  0.520 &          0.36 &           0.39 & -0.03 &                   0.306 \\
          verb &  4352186.5 &  0.007 &  0.518 &         12.57 &          12.28 &  0.29 &                   0.606 \\
   affiliation &  4354677.0 &  0.008 &  0.518 &          1.86 &           1.81 &  0.05 &                   0.671 \\
       achieve &  4353813.5 &  0.008 &  0.518 &          1.65 &           1.70 & -0.05 &                   0.648 \\
           bio &  4350804.0 &  0.007 &  0.518 &          1.29 &           1.04 &  0.25 &                   0.571 \\
          feel &  4360879.0 &  0.010 &  0.517 &          0.32 &           0.29 &  0.03 &                   0.835 \\
       article &  4366549.5 &  0.012 &  0.517 &          8.54 &           8.42 &  0.12 &                   1.074 \\
      informal &  4368808.5 &  0.013 &  0.516 &          0.38 &           0.38 &  0.00 &                   1.152 \\
         shehe &  4373985.5 &  0.017 &  0.516 &          1.89 &           1.77 &  0.12 &                   1.420 \\
            we &  4393997.0 &  0.033 &  0.514 &          0.59 &           0.63 & -0.04 &                   2.829 \\
        sexual &  4407971.5 &  0.024 &  0.512 &          0.11 &           0.13 & -0.02 &                   2.072 \\
        Period &  4411906.5 &  0.058 &  0.512 &          5.36 &           5.35 &  0.01 &                   5.019 \\
        number &  4406451.0 &  0.050 &  0.512 &          2.05 &           2.36 & -0.31 &                   4.258 \\
         swear &  4411774.5 &  0.008 &  0.512 &          0.03 &           0.03 &  0.00 &                   0.656 \\
        nonflu &  4405019.5 &  0.031 &  0.512 &          0.08 &           0.08 &  0.00 &                   2.694 \\
   focusfuture &  4421046.5 &  0.076 &  0.511 &          1.00 &           1.01 & -0.01 &                   6.530 \\
         cause &  4422613.0 &  0.079 &  0.511 &          1.60 &           1.62 & -0.02 &                   6.820 \\
         Colon &  4426143.5 &  0.086 &  0.510 &          0.40 &           0.36 &  0.04 &                   7.421 \\
       compare &  4457029.5 &  0.184 &  0.507 &          2.65 &           2.66 & -0.01 &                  15.855 \\
        female &  4451589.0 &  0.156 &  0.507 &          0.69 &           0.63 &  0.06 &                  13.447 \\
        health &  4456542.5 &  0.181 &  0.507 &          0.65 &           0.47 &  0.18 &                  15.608 \\
        friend &  4467566.0 &  0.220 &  0.506 &          0.16 &           0.17 & -0.01 &                  18.889 \\
       Parenth &  4464664.0 &  0.212 &  0.506 &          0.56 &           0.52 &  0.04 &                  18.200 \\
        Exclam &  4460400.0 &  0.098 &  0.506 &          0.04 &           0.05 & -0.01 &                   8.416 \\
         ipron &  4476870.0 &  0.273 &  0.505 &          4.74 &           4.71 &  0.03 &                  23.461 \\
          male &  4498125.5 &  0.386 &  0.502 &          1.63 &           1.63 &  0.00 &                  33.224 \\
        family &  4507666.5 &  0.438 &  0.501 &          0.27 &           0.29 & -0.02 &                  37.705 \\
           you &  4510550.5 &  0.458 &  0.501 &          0.48 &           0.48 &  0.00 &                  39.401 \\
       insight &  4508382.5 &  0.446 &  0.501 &          1.96 &           1.95 &  0.01 &                  38.328 \\
        filler &  4509865.0 &  0.376 &  0.501 &          0.01 &           0.01 &  0.00 &                  32.359 \\
       pronoun &  4507193.5 &  0.439 &  0.501 &          9.27 &           9.08 &  0.19 &                  37.729 \\
         ppron &  4517453.5 &  0.499 &  0.500 &          4.53 &           4.37 &  0.16 &                  42.933 \\
          body &  4515469.5 &  0.487 &  0.500 &          0.30 &           0.28 &  0.02 &                  41.912 \\
\bottomrule
\end{longtable}

\end{document}